\pdfoutput=1
\documentclass[aps,prl,10pt,twocolumn,floats,floatfix,showkeys,showpacs,superscriptaddress]{revtex4}
\usepackage{graphicx,amssymb,amsmath}
\usepackage[latin1]{inputenc}
\usepackage{color}
\begin{document}

\title{Resiliently evolving supply-demand networks}
\author{Nicol{\'a}s Rubido}
\email{n.rubido.obrer@abdn.ac.uk}
\affiliation{Institute for Complex Systems and Mathematical Biology, University of Aberdeen,
	     King's College, AB24 3UE Aberdeen, UK}
\affiliation{Instituto de F\'{i}sica, Facultad de Ciencias, Universidad de la Rep\'{u}blica,
	     Igu\'{a} 4225, Montevideo, 11200, Uruguay}
\author{Celso Grebogi}
\affiliation{Institute for Complex Systems and Mathematical Biology, University of Aberdeen,
	     King's College, AB24 3UE Aberdeen, UK}
\author{Murilo S. Baptista}
\affiliation{Institute for Complex Systems and Mathematical Biology, University of Aberdeen,
	     King's College, AB24 3UE Aberdeen, UK}
%
%
\date{\today}
\begin{abstract}
The ability to design a transport network such that commodities are brought from suppliers
to consumers in a steady, optimal, and stable way is of great importance for nowadays
distribution systems. In this Letter, by using the circuit laws of Kirchhoff and Ohm, we
provide the exact capacities of the edges that an optimal supply-demand network should have
to operate stably under perturbations. The perturbations we consider are the evolution of
the connecting topology, the decentralisation of hub sources or sinks, and the intermittence
of suppliers/consumers characteristics. We analyse these conditions and the impact of our
results, both on the current UK power-grid structure and on numerically generated evolving
archetypal network topologies.
\end{abstract}

\maketitle
Networks are ubiquitous in nature and man-made systems. Power and gas networks bring light
and heating to our homes, telecommunication networks allow us to be entertained and to
browse for information, and distribution networks allow manufacturers to supply food-stock
and other products to the demand chain. In all of these cases a basic problem needs to be
addressed: how to create a steady, optimal, and stable transport of commodities across such
supply-demand networks.

We understand that a supply-demand network is \emph{stable} when the system is not vulnerable
to modifications in the network's connectivity, switch from hub suppliers to decentralised
smaller producers, or changes in the location of suppliers and/or consumers that may cause
over-load failures to occur. 
It is \emph{optimal} when the transport is done such that the cost is minimum, e.g.,
minimising the energy consumption.

Our problem is part of what flow network theory tries to decipher \cite{Kirchhoff,Bollobas,
Ahuja,Zhou}, a theory that roots back to Kirchhoff \cite{Kirchhoff}, answering what are the
current flows in each edge of an electrical circuit as a set of voltages are applied to some
nodes. The solution is then achieved by solving \emph{Kirchhoff's equations}. It is related
to the probability that a random walker starts at the source and finishes at the sink
\cite{FanChung} and to first-passage times at each node \cite{Randall}. In order to model a
supply-demand network, we solve the inverse problem. Namely, we calculate voltages (loads)
using a conservative minimal cost transport system when input/output currents (flows) are
given. This means loads are carried optimally from the source (supplier) to the sink
(consumer) without losses. It is related to finding shortest-paths and community structures
on weighted networks \cite{Newman1,Newman2}.

In particular, Kirchhoff's flow network model is used to express electrical flow in circuits
\cite{Rubido}, but also to establish systems ecology quantities relationship \cite{Brown},
biologically inspired steady-state's transport systems \cite{Katifori}, and fractures in
materials \cite{Batrouni,Pinheiro}. Although the relationship between flows and loads in
these models is restricted to be linear and conservative, the complexity in the mathematical
treatment of the equations due to the topology structure is still demanding. Thus, most flow
network solutions are based on optimisation schemes \cite{Ahuja}. The results are complex and
not easy to relate to other relevant network centrality measures. Moreover, if the network
evolves in time (the connecting topology changes with nodes and/or edges appearing and/or
disappearing), then predictions, controlling cascade of failures, and analytical solutions
are scarcer \cite{Havlin1,Havlin2,Lai1,Lai2}.

In this Letter, we provide analytical expressions for the edge capacities that a steady
optimal supply-demand network should have to operate stably under perturbations by using
Kirchhoff flow network model. The perturbations we consider constitute some possible
evolution factors that supply-demand networks are subjected to, such as, switching from hub
suppliers to multiple smaller producers, intermittent supplying and consuming nodes, and
either node and edge additions or removals.

We apply our edge capacity analytical results both to the current UK power-grid and to
numerically generated evolving archetypal network topologies. We discuss the design of a
modern steady-state stable power-grid system and we find that most topology modifications
have a power-law behaviour of the analytically derived edge capacities as node or edges are
added to the network. Our results and conclusions are general and applicable to any other
system that is modelled by Kirchhoff flow network model in its steady-state. Furthermore,
they are related to standard network characteristics and allow to predict cascade of
failures due to over-loads.

To achieve an analytical solution of our problem, we initially assume that the network
structure, the location of the supplier(s) and consumer(s), and the amount of commodities
produced and consumed are known, but loads and flows in every edge need to be calculated.
Furthermore, we let loads to be linearly related to the flows by
\begin{equation}
  l_{ij}^{(st)} = R_{ij}\,f_{ij}^{(st)}\,.
 \label{eq_load_flow_rel}
\end{equation}
The left hand side of Eq.~(\ref{eq_load_flow_rel}) is the load being transferred across the
edge connecting nodes $i$ and $j$ of the network given a source located at node $s$ and a sink
located at node $t$. The extension to many sources and sinks adds more upper indexes to the
equation (see Supplementary Material). In our model, $l_{ij}^{(st)}$ represents the voltage 
difference between two points in an electric circuit, where a current enters the circuit at
node $s$ and leaves at node $t$ \cite{Rubido}. 
The right hand side of Eq.~(\ref{eq_load_flow_rel}) is composed of
a proportionality factor $R_{ij}$, depending on the structure of the edge, and the unknown
flow $f_{ij}^{(st)}$, which the edge has for that particular location of the supply-demand
nodes. Physically, it is the edge's resistance times the electrical current. 
For example, in systems ecology,
Eq.~(\ref{eq_load_flow_rel}) relates the storage quantities $Q$ with the outflows $J$ via
time constants $T$ \cite{Brown}.

Because we assume the model to be conservative, the net flow at any node $i$ in the network
is null ($\sum_{j=1}^N f_{ij}^{(st)} = 0$), with the exception of the source (whose net flow
is represented by $I$) and of the sink (whose net flow is represented by $-I$) nodes. This
guarantees that flows are carried optimally from the source(s) to the sink(s). Then, the net
flow at node $i$ is
\begin{equation}
  \sum_{j=1}^N f_{ij}^{(st)} = I\,\left(\delta_{is} - \delta_{it}\right) = \sum_{j = 1}^N
   W_{ij}\left( V_i^{(st)} - V_j^{(st)} \right)\,,
 \label{eq_init_flows}
\end{equation}
where $V_i^{(st)}$ is the voltage potential at node $i$ for the particular $s$-$t$ pair
and $W_{ij} = A_{ij}/R_{ij}$ is the matrix representation of the network structure. $A_{ij}
= A_{ji}$ is the adjacency matrix entries, that are either $1$ (nodes $i$ and $j$ are
connected to each other) or $0$ ($i$ and $j$ are not connected), and $R_{ij}$ is the edge
resistance.

We also make use of the equivalent resistance $\rho_{ij}$, which is a network structure
characteristic \cite{Cserti,Wu,Arpita}. It is found analytically from the weighted Laplacian
matrix $\mathbf{G}$ eigenvalues and eigenvectors ($\mathbf{G} = \mathbf{D} - \mathbf{W}$,
with $D_{ij} = \delta_{ij}\,k_j$, and $k_i = \sum_{i=1}^N W_{ij}$, $N$ being the number of
nodes in the network). This quantity allows to express any connected arbitrary topology
(defined with or without weighted edges) to an effective weighted complete network.

Our first analytical result is the edge capacity $C_{ij}^{(st)}$ that a supply-demand network
must have in order to operate stably avoiding over-loads is such that \mbox{$\left|l_{ij}^{
(st)}\right|\leq C_{ij}^{(st)}$}. We obtain that \emph{$C_{ij}^{(st)}$ is a function of the
equivalent resistance of the edge and is proportional to the total amount of commodities per
unit of time that are produced by the suppliers} (the total input $I$). Moreover, the value
of $C_{ij}^{(st)}$ we find is independent of where the suppliers and consumers are located
within the network (namely, $C_{ij}^{(st)} = C_{ij}$). Specifically, \emph{the exact value
an edge capacity} must have is
\begin{equation}
  C_{ij} \equiv I\,\rho_{ij}\,,
 \label{eq_edge_capacity}
\end{equation}
where the functional $\rho_{ij}(\mathbf{W})$ is the equivalent resistance between nodes $i$
and $j$ \cite{Rubido}. \emph{The edge capacities in} Eq.~(\ref{eq_edge_capacity}) \emph{
quantify the value that each existing edge of the supply-demand network must have to secure
a steady-state stable distribution, regardless of the location of the producer and consumer
and regardless if, instead of a single supplier and consumer, there are many with arbitrary
spatial distributions} [the demonstration of Eq.~(\ref{eq_edge_capacity}) from
Eq.~(\ref{eq_init_flows}) is given in the Supplementary Material]. To derive the exact values
for node capacity $C_i$, we perform a summation over all edge capacities $C_i = \sum_{j = 1}^N
C_{ij}$, which is feasible because our model of flow network is conservative. Hence, the node
and edge capacities are not independent quantities.

Cascade of failures on networks are often studied by analysing how attacks and/or over-loads
occur when a load surpass the node capacity. Such a node capacity is conjectured to have,
with some tuning parameters, a linear relationship with the initial load distribution \cite{
Lai1,Lai2}. This assumption allows to draw conclusions on how the network structure should be
designed to avoid failures due to over-loads. Here, \emph{we show that the capacity-load
relationship is given by $\rho_{ij}$} [Eq.~(\ref{eq_edge_capacity})], \emph{and it is derived
from finding the edge's maximum loads for any of the discussed network evolution processes}.
In the cases that the physical edge capacity is pre-assigned, such as in a fuse network (a
model that explains fractures in materials \cite{Batrouni,Pinheiro}), then
Eq.~(\ref{eq_edge_capacity}) predicts exactly which edges will over-load due to the
perturbations. This can still aid in the prevention of cascade of failures as it detects the
vulnerable edges exactly.

The second analytical result we find is that \emph{all $C_{ij}$ defined by}
Eq.~(\ref{eq_edge_capacity}) \emph{are bounded by the inverse of the largest} ($\lambda_{
N-1}\!\left(\mathbf{G}\right)$) \emph{and the smallest non-zero} ($\lambda_1\!\left(\mathbf{G}
\right)$, also known as spectral gap) \emph{weighted Laplacian matrix $\,G_{ij}$ eigenvalues}
\cite{Zhang,Das}. In particular, we find that
\begin{equation}
  \frac{2I}{\lambda_{N-1}}\left( 1 - \delta_{ij} \right) \leq C_{ij} \leq
   \frac{2I}{\lambda_{1}}\left( 1 - \delta_{ij} \right)\,.
 \label{eq_capacity_bounds}
\end{equation}
Moreover, these eigenvalues are related to the minimal and maximal degrees of the network,
thus, providing a way to modify the network topology (adding or removing nodes and edges)
and keep the capacity values bounded by considering simple rules from minimal information
about the structure [the derivation of Eq.~(\ref{eq_capacity_bounds}) and the relationship
of these bounds to the node degrees is provided in the Supplementary Material].

As a practical proof of concept and a way to illustrate these analytical results, we apply
them both to the UK power-grid structure \cite{UK_grid_data} and numerically generated random
\cite{Erdos} and small-world \cite{Strogatz} topologies. In these frameworks, we discuss the
following perturbations: (\emph{i}) \emph{power generator decentralisation} (changing from
centralised high-power generators to distributed smaller generators), (\emph{ii})
\emph{source-sink intermittency} (the inclusion of suppliers and consumers, such as renewable
sources, storage systems, and electric cars, all possibly changing locations within the
network), and \emph{(iii) connectivity modifications}. These are some of the most important
perturbations that modern power-grid systems are having to deal with. We address cases
(\emph{i}) and (\emph{ii}) using the real UK power-grid structure. For the case (\emph{iii}),
we focus the analysis on how the connectivity modifications affect the edge's capacity by
deriving them for numerically generated networks.

\begin{figure}[htbp]
 \begin{center}
  \includegraphics[width=1.0\columnwidth]{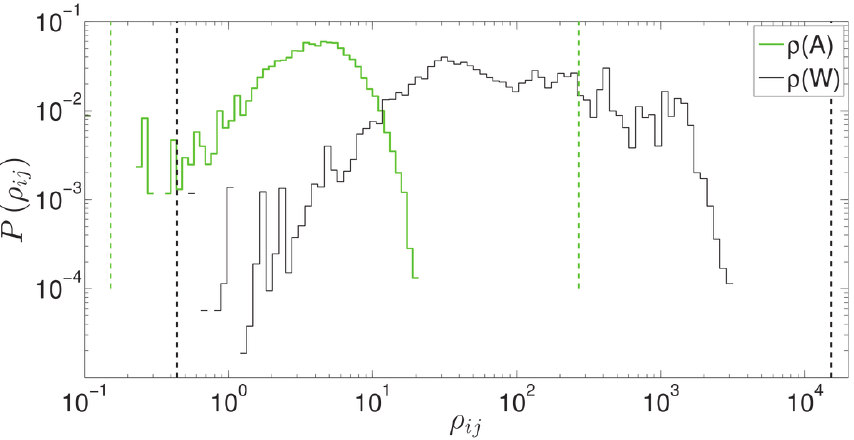}
 \end{center} \vspace{-1pc}
  \caption{The plot shows the probability density functions of the equivalent resistance
  $\rho_{ij}$ (stairs-like lines) for the power-grid adjacency (green online) and weighted
  structure (black), and their respective bounds (vertical dashed lines are the bounds of
  Eq.~(\ref{eq_capacity_bounds})). The units of $\rho_{ij}$ are in MVA (as the power-line
  resistances found from \cite{UK_grid_data}) and include all three UK's major transmission
  companies (SHETL, SPT, and NGET).}
 \label{fig_UK_power_grid_ER}
\end{figure}

Since the capacity is linearly related to $\rho_{ij}$ [Eq.~(\ref{eq_edge_capacity})], to
obtain the influence of the topology on the loads we first calculate the UK power-grid
$\rho_{ij}$ by assuming $R_{ij} = 1$, i.e., $W_{ij} = A_{ij}$. The resultant set of
$\rho_{ij}$ is represented by $\rho(A)$ in Fig.~\ref{fig_UK_power_grid_ER}. We then calculate
$\rho_{ij}$ considering the edge's resistance in MVA units (mega Volt-Ampere)
\cite{UK_grid_data}, i.e., $W_{ij} = A_{ij}/R_{ij}$. This set is represented by $\rho(W)$ in
Fig.~\ref{fig_UK_power_grid_ER}. The bounds in terms of the maximum and minimum eigenvalues
[Eq.~(\ref{eq_capacity_bounds})] are shown by the vertical dashed lines in each case.

From an engineering point of view, the steady-state stability of a power network is the
capability of the system to maintain the power transmitted between any two nodes below
the edge capacity (namely, the maximal load that the connecting power line can handle) when
perturbations are applied to the network. Figure~\ref{fig_UK_power_grid_ER} shows that,
if we neglect the reactance and inductance characteristics of the power lines and we model
the power-grid by a conservative linear flow network model such as Eq.~(\ref{eq_load_flow_rel}),
\emph{then assigning edge capacities to the edges in the power-grid drawn from the $\rho$
distribution guarantees the steady-state stability of the system}. Such a power-grid system
is resilient to changing from hub generators to distributed sources [case (\emph{i})] or
having intermittent sources and sinks [case (\emph{ii})].

In the case (\emph{i}), a single source is substituted by multiple sources while maintaining
the same inflow of power. Namely, $I$ is kept constant in the transformation while the network
structure does not change. In such a case, our first analytical result for $C_{ij}$
[Eq.~(\ref{eq_edge_capacity})] predicts that the new maximum loads are always less than the
capacity value $\left| l_{ij}^{(s_1,s_2,\ldots,t_1,t_2,\ldots)} \right|<\left| l_{ij}^{(st)}
\right| \leq I\,\rho_{ij}$. In the case (\emph{ii}), given the intermittency property of
renewable sources and electric car power stations, the supply-demand network behaves as if
sources and sinks keep changing locations with time. In other words, the system explores
various configurations of the many source-sink problem for a fixed structure at different
times. Equation~(\ref{eq_edge_capacity}) describes the edge capacities for all edges that
ensures no over-load will happen.

On the contrary, \emph{the modifications to the connecting topology} [case (\emph{iii})]
\emph{change the value of $\rho_{ij}$, which consequently, changes the edge capacities and
redistributes the flows}. In this case, in order to draw conclusions about edge capacities
for a power-grid such as the UK, one needs a dynamic picture of the network topology as it
evolves. We find that the minimum and maximum eigenvalues [Eq.~(\ref{eq_capacity_bounds})]
must be kept fixed to design edge capacities with fixed margins as the supply-demand network
topology changes. In this sense, the changes in the topology are contained within the bounds
of the initial edge capacities. Thus, resilience is inforced by using the upper bound of
Eq.~(\ref{eq_capacity_bounds}) for every edge. 
In order to keep the minimal and maximum eigenvalues fixed, we find that it is enough to fix
the minimum and maximum node degrees of the network (Supplementary Material).

In general, \emph{we find that if a supply-demand network needs to be designed with similar
edge capacities, then the graph has to be set such that it resembles as much as possible a
complete graph} (which has all its non-null eigenvalues equal to the node degrees). This
narrows the values that $\rho_{ij}$ can take. On the other hand, if the range of $C_{ij}$
values sought needs to be as broad as possible, then the network should be designed in such
a way it include some nodes with higher degrees (this increases the largest eigenvalue, thus,
diminishing the lower bound for the edge capacity) and, well-defined communities (which
lowers the magnitude of the spectral gap, hence, increasing the upper bound for the edge
capacity) or nodes with lower degrees.

\begin{figure}[htbp]
  \begin{center}
  \begin{minipage}{10pc}
   \textbf{(a)}\\
   \includegraphics[width=10pc]{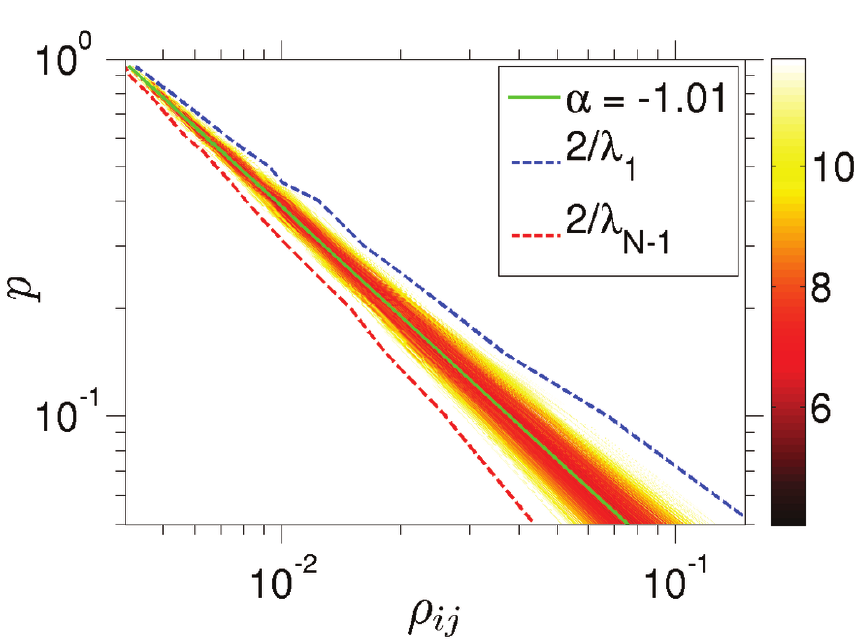}
  \end{minipage}
  \begin{minipage}{10pc}
   \textbf{(b)}\\
   \includegraphics[width=10pc]{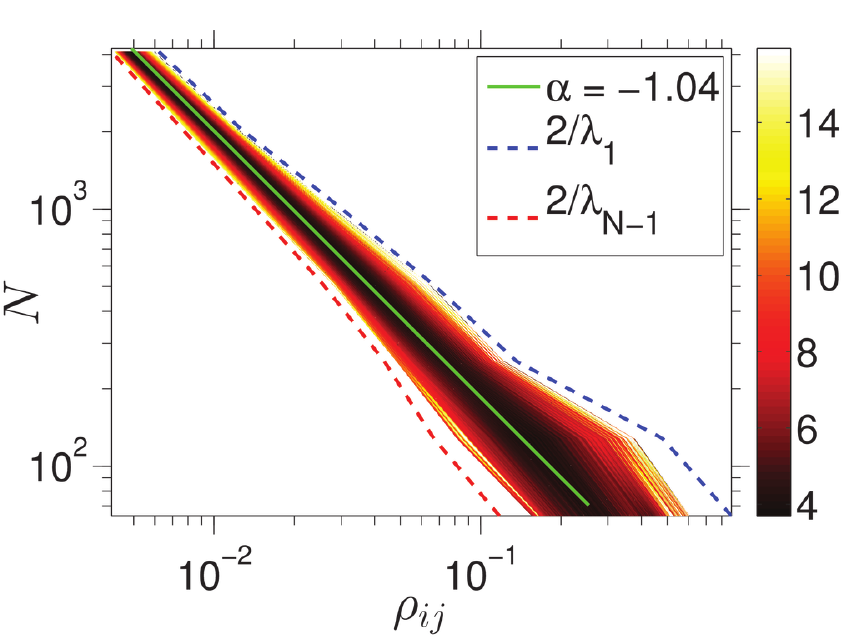}
  \end{minipage}
 \end{center} \vspace{-1pc}
  \caption{
  Edge [panel {\bf (a)}] and node [panel {\bf (b)}] addition protocols effect on the PDF of
  the $\rho_{ij}$ of random networks. Panel {\bf (a)} simulations start from a ring graph
  of $N = 2^9$ nodes. Then, edges linking two disjoint nodes are added with probability $p
  \in[0,1]$. The node addition in panel {\bf (b)} is done by growing the ring graph from $N
  = 2^6$ to $2^{12}$. Then, nodes are linked with probability $p=10^{-1}$. 
  The colour scale corresponds to the logarithm of the PDF's values. The analytical
  bounds of Eq.~(\ref{eq_capacity_bounds}) are shown with dashed lines (see insets).}
 \label{fig_equiv_resist_RN}
\end{figure}

\begin{figure}[htbp]
 \begin{center}
  \begin{minipage}{10pc}
   \textbf{(a)}\\
   \includegraphics[width=10pc]{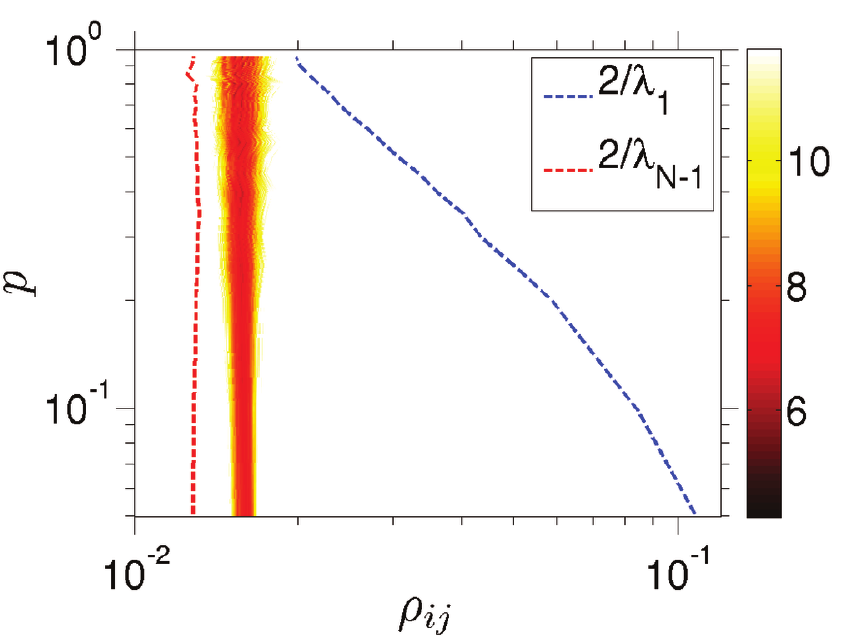}
  \end{minipage}
  \begin{minipage}{10pc}
   \textbf{(b)}\\
   \includegraphics[width=10pc]{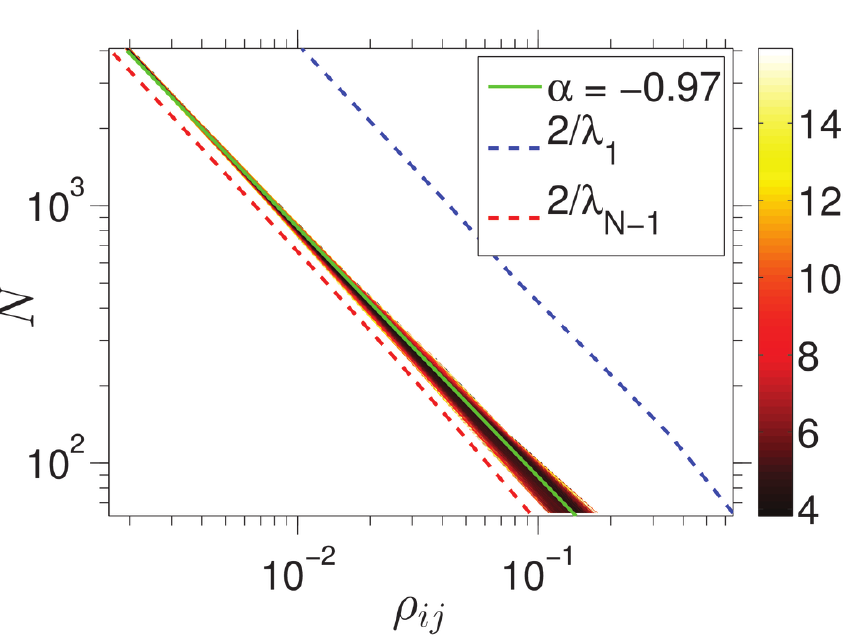}
  \end{minipage}
 \end{center} \vspace{-1pc}
  \caption{
  Edge rewiring [panel {\bf (a)}] and node addition [panel {\bf (b)}] protocols effect on the
  PDF of the $\rho_{ij}$ of small-world networks. Panel {\bf (a)} simulations are carried out
  for a regular network of $N = 2^9$ nodes with node degree $k = N/4$ for all nodes. Then,
  edges are rewired with probability $p\in[0,1]$. The node addition in panel {\bf (b)} is
  done by growing the regular graph from $N = 2^6$ to $2^{12}$. Then, edges are rewired
  with probability $p = 10^{-1}$. 
  Colour scale and lines follow the same criteria as in Fig.~\ref{fig_equiv_resist_RN}.}
 \label{fig_equiv_resist_SW}
\end{figure}

We particularise now the analysis of the effect of connectivity modifications to the
$\rho_{ij}$ probability distribution function (PDF) for two types of numerically generated
networks: random (Fig.~\ref{fig_equiv_resist_RN}) \cite{Erdos} and small-world
(Fig.~\ref{fig_equiv_resist_SW}) \cite{Strogatz}. In both cases, two growth protocols are
carried out. The effect of these protocols on the $\rho_{ij}$ PDFs are shown in
Figs.~\ref{fig_equiv_resist_RN} and \ref{fig_equiv_resist_SW}.

For \emph{random networks} (RN), the first protocol for RN keeps the number of nodes fixed
but increases the number of edges (hence, increasing the density of edges). The second
protocol adds nodes maintaining the average number of edges fixed (thus, decreasing the
density of edges). For \emph{small-world networks} (SW), the first protocol rewires the
existing edges in a regular graph. This means the number of nodes, edges, and density of
edges, are fixed. The second protocol increases the number of nodes and edges but keeps
the average edge density constant.

RN and SW networks exhibit power-law behaviour of the $\rho_{ij}$ PDFs for both growth
protocols, with the exception of the edge rewiring protocol for fixed number of nodes in SW
networks [panel {\bf (a)} in Fig.~\ref{fig_equiv_resist_SW}]. In other words, most growth
processes lead to a power-law distribution of the edge capacities as a function of the
control parameter (either $p$ or $N$), as it also happens in scale-free networks
\cite{Havlin1}. This is very advantageous when designing an invariant flow distribution for
an evolving supply-demand network. Moreover, we find that at every step of the growth
process the edge capacity distribution is mainly given by the behaviour of the most probable
$\bar{\rho}$ value. Thus, its evolved magnitude can be predicted from the power-law
exponents at any step. The evolution of $\bar{\rho}$ 
is derived from
\begin{equation}
  \bar{\rho}(r) \simeq e^{-\beta/\alpha}\,r^{1/\alpha}\,,
 \label{eq_scaling_law}
\end{equation}
where $\alpha < 0$ and $\beta$ are the scaling exponents [$\log(r) \simeq \alpha\,\log
\left(\bar{\rho}\right) + \beta$] and the protocol control parameter $r$ is either $p$ or
$N$. Furthermore, as can be seen from the insets of these figures, we find that $\alpha\sim
-1$ for all the power-law cases.

We interpret the power-law behaviour in the following way. Any addition of edges results
in a decrease of the equivalent resistance between nodes. This is a consequence of having
shorter paths between nodes, namely, more connecting edges. This is why the edge addition
protocol for RN and the node addition protocol for SW result in a power-law PDF evolution.
The node addition protocol for RN keeps an average number of edges fixed, though it diminishes
the edge density at every stage, thus, creating a power-law evolution of the edge's
$\rho_{ij}$ due to the existence of some short and long paths balance. \emph{When the
connectivity modification protocol fixes the number of nodes, edges, and the edge density}
[such as the protocol in panel {\bf (a)} of Fig.~\ref{fig_equiv_resist_SW}], \emph{then the
$\rho_{ij}$'s PDF remains invariant}.

Any real supply-demand network operates under unpredictable fluctuations, such as, the switch
from hub sources to distributed smaller producer [case (\emph{i})], the change in the
location of suppliers and consumers [case (\emph{ii})], topology intended modifications
[case (\emph{iii})], or directed attacks. Using Eq.~(\ref{eq_edge_capacity}), we provide a
robust edge capacity that is not surpassed in either case (\emph{i}) or case (\emph{ii}).
In the case (\emph{iii}), we find that most topology evolution factors cause the edge
capacity to evolve in a power-law behaviour. 

In the cases where the knowledge of the full network structure is missing, we also obtain
manageable bounds for the exact capacity values in terms of minimal information of the
network structure, e.g., eigenvalues of the weighted Laplacian matrix
[Eq.~(\ref{eq_capacity_bounds})] and minimum/maximum degrees (see Supplementary Material).
Our margins give simple engineering strategies for modifying the network's topology while
bounding the capacities and maintaining a stable distribution [case (\emph{iii})]. 


To summarise, by analytically providing exact edge capacities values (plus bounds) of
conservative linear flow problems, we are able to show how to design resiliently evolving
supply-demand networks.

Authors acknowledge the Scottish University Physics Alliance (SUPA).




\begin{thebibliography}{10}
%
%
 \bibitem{Kirchhoff} G. Kirchhoff, {\it Ueber die Aufl{\"o}sung der Gleichungen, auf welche man bei der Untersuchung der linearen Vertheilung galvanischer Str{\"o}me gef{\"u}hrt wird}, (Wiley Online Library, Annalen der Physik, 1847).
 \bibitem{Bollobas} B. Bollob{\'a}s, {\it Modern Graph Theory} (Springer-Verlag, New York, 1998).
 \bibitem{Ahuja} R. K. Ahuja, T. L. Magnanti, and J. B. Orlin, {\it Network Flows: Theory, Algorithms, and Applications} (Prentice-Hall, Ch. 1 and 3, 1993).
 \bibitem{Zhou} Y.-H. Chen, B.-H. Wang, L.-C. Zhao, C. Zhou, and T. Zhou, Phys. Rev. E {\bf 81}, 066105 (2010).
%
 \bibitem{FanChung} F. R. K. Chung, {\it Spectral Graph Theory} (Am. Math. Soc., CBMS 92, Ch. 1, 3, 4, and 7 1997).
 \bibitem{Randall} D. Randall, Comp. in Sci. and Eng. {\bf 6}, 1521-9615 (2006).
%
 \bibitem{Newman1} M. E. J. Newman and M. Girvan, Phys. Rev. E {\bf 69}, 026113 (2004).
 \bibitem{Newman2} M. E. J. Newman, Eur. Phys. J. B {\bf 38}, 321-330 (2004).
%
 \bibitem{Rubido} N. Rubido, C. Grebogi, and M. S. Baptista, EPL {\bf 101}, 68001 (2013).
 \bibitem{Brown} M. T. Brown, Ecol. Mod. {\bf 178}, 83-100 (2004).
 \bibitem{Katifori} E. Katifori, G. J. Szollosi, and M. O. Magnasco, Phys. Rev. Lett. {\bf 104}, 048704 (2010).
 \bibitem{Batrouni} G. G. Batrouni, and A. Hansen, Phys. Rev. Lett. {\bf 80}(2), 325(4) (1998).
 \bibitem{Pinheiro} C. F. S. Pinheiro and A. T. Bernarde, Phys. Rev. E {\bf 72}, 046709 (2005).
%
 \bibitem{Havlin1} E. L\'opez, S. V. Buldyrev, S. Havlin, and H. E. Stanley, Phys. Rev. Lett. {\bf 94}, 248701 (2005).
 \bibitem{Havlin2} S. Carmi, Z. Wu, S. Havlin, and H. E. Stanley, EPL {\bf 84}, 28005 (2008).
 \bibitem{Lai1} R. Yang, W.-X. Wang, Y.-C. Lai, and G. Chen, Phys. Rev. E {\bf 79}, 026112 (2009).
 \bibitem{Lai2} W.-X. Wang and Y.-C. Lai, Phys. Rev. E {\bf 80}, 036109 (2009).
%
 \bibitem{Cserti} J. Cserti, Am. J. Phys. {\bf 68} (10), 896-906 (2000).
 \bibitem{Wu} F. Y. Wu, J. Phys. A: Math. Gen. {\bf 37}, 6653-6673 (2004).
 \bibitem{Arpita} A. Ghosh, S. Boyd, and A. Saberi, SIAM Rev. {\bf 50}(1), 37-66 (2008).
 \bibitem{Zhang} H. Chen and F. Zhang, Disc. App. Math. {\bf 155}, 654-661 (2007).
 \bibitem{Das} K. Ch. Das, A. D. G\"ung\"or, and A. S. Cevic, MATCH Commun. Math. Comput. Chem. {\bf 67}, 541-556 (2012).
%
 \bibitem{UK_grid_data} www.nationalgrid.com/uk/Electricity/SYS/current
%
%
 \bibitem{Erdos} P. Erdos and A. Renyi, Publ. Math. {\bf 6}, 290 (1959). 
 \bibitem{Strogatz} D. J. Watts and S. H. Strogatz, Nature {\bf 393}, 440-442 (1998).
%
%
\end{thebibliography}
\end{document}